\documentstyle[preprint,aps]{revtex}

\def\journal#1, #2, #3, 1#4#5#6{
    {\sl #1~}{\bf #2}, #3 (1#4#5#6)}

\def\prd{\journal Phys. Rev. D, }
\def\prl{\journal Phys. Rev. Lett., }

\def\cmp{\journal Comm. Math. Phys., }
\def\np{\journal Nucl. Phys., }
\def\pl{\journal Phys. Lett., }

\def\CC{{\cal C}}
\def\RR{{\cal R}}

\def\MM{{\cal M}}
\def\SS{{\cal S}}

\begin{document}
\preprint{UdeM-GPP-TH-95-22}

\title{Remarks on gauge vortex scattering}

\author{R. MacKenzie}
\address{Laboratoire de Physique Nucl\'eaire, Universit\'e de
Montr\'eal C.P. 6128, succ. centreville, Montr\'eal, Qu\'ebec, Canada, H3C 3J7}
\maketitle

\begin{abstract}
\widetext

In the abelian Higgs model, among other situations, it has recently been
realized that the head-on scattering of $n$ solitons distributed
symmetrically around the point of scattering is by an angle $\pi/n$,
independant of various details of the scattering. In this note, it is first
observed that this result is in fact not entirely surprising: the above is
one of only two possible outcomes. Then, a generalization of an argument
given by Ruback for the case of two gauge theory vortices in the Bogomol'nyi
limit is used to show that in the geodesic approximation the above result
follows from purely geometric considerations.

\narrowtext \end{abstract}
\bigskip\bigskip
\centerline{hep-th/9503044}
\newpage

The scattering of topological solitons exhibits many interesting and
surprising features in a wide variety of situations.\footnote{We use the
``physicist's definition'' of soliton; we are not concerned here with
integrability, elasticity of scattering, etc.} For instance, the scattering
of monopoles\cite{mono}, of vortices\cite{vort}, of skyrmions\cite{skyr}
and of baby skyrmions\cite{baby}  have many common aspects, such as the
fact that, at low energies, head-on scattering in many situations is at
ninety degrees. This has been seen both analytically and numerically.
Generalizations have been studied wherein the scattering of several
solitons has been considered\cite{nsk,nmon,nvort,nbaby}. Typically, the
initial condition chosen is the most symmetric one possible: one considers
any number $n$ of solitons directed with equal energies towards the origin
with relative angular separations $2\pi/n$, starting at equal distances,
with zero impact parameter (which could be described as an $n$-soliton
head-on collision). The outcome of such scatterings is rather beautiful:
the radii along which the solitons leave after scattering are rotated
relative to the incident angles by an angle of $\pi/n$.  Thus, for
instance, if four particles are incident along the positive and negative
$x$ and $y$ axes, after scattering, the particles will leave along the four
axes $x=\pm y$, while if three particles are incident with polar angles
$\pi/3$, $\pi$ and $5\pi/3$, they scatter along radii of polar angles $0$,
$2\pi/3$ and $4\pi/3$.

While at first sight surprising, this result {\sl almost} follows directly
from purely symmetric considerations. Let us consider the possible outcomes
of such a 2+1-dimensional ``experiment'' with $n$ solitons incident at
angular separations of $2\pi/n$. First, due to the very nature of solitons,
the winding number cannot change. If no bound states of solitons exist,
there must necessarily be $n$ distinct solitons after the scattering. Given
that the initial configuration is symmetric under rotation by $2\pi/n$ and
under reflections (that is, the symmetry group of a regular $n$-gon), the
only possible final states which also have this symmetry are one where the
solitons leave along the same radii as the incident ones, or one where the
final radii are situated midway between the incident ones. It is convenient
to adopt a slightly {\sl unconventional} definition of scattering angle,
namely, as the angle between an incoming radial direction and the nearest
outgoing radial direction (rather than measuring scattering relative to the
straight-through direction). Then the first possibility corresponds to zero
scattering angle while the second corresponds to a scattering of
$\pi/n$.\footnote{The reason for adopting this unconventional definition is
that with the traditional definition the cases of even or odd number of
solitons must be treated separately: the first possibility corresponds to
traditional scattering angle $0$ or $\pi/n$ for $n$ even or odd,
respectively, while the second corresponds to traditional scattering angle
$\pi/n$ or $0$ for $n$ even or odd, respectively. Note that, since the
solitons are identical, scattering is in fact only defined modulo
$2\pi/n$.}

In what follows, an argument will be given which shows that in the case of
slow-moving gauge theory vortices in the Bobomol'nyi limit\cite{bog},
scattering is necessarily at an angle of $\pi/n$. The argument is a
straightforward generalization of the  result of Ruback\cite{ruback}, who
considered the scattering of two vortices in this limit. Following similar
reasoning of Manton\cite{manton}, Ruback first noted that in this limit
there are no static  forces between vortices: essentially, the force
mediated by the gauge particle is cancelled by that mediated by the Higgs
particle. This implies that the motion of slow-moving vortices will be
along ``troughs'' in field space which are the static $n$-vortex
configurations parameterized by the positions of the zeroes of the Higgs
field. This space of configurations is a $2n$-dimensional subspace of the
whole field space and is the ``moduli space'' of $n$ vortices,
$\MM_n$\cite{moduli}. The motion of vortices at low energy is, in fact,
geodesic motion with metric determined by the original field theory action,
applied to configurations in $\MM_n$.

In the case of two vortices, $\MM_2$ clearly separates into two two-dimensional
subspaces $\CC$ and $\RR$, representing center-of-mass and relative motion.
It is physically obvious that motion in $\CC$ is trivial due to
translational invariance; mathematically, this invariance implies that the
induced metric on $\MM_2$ is in fact block-diagonal, there being no mixed term
between $\CC$ and $\RR$. Thus, one can deduce the dynamics of two vortices
by studying the induced metric in $\RR$ only.

In the case of $n$ vortices, $\MM_n$ is somewhat more complicated. Two of
the $2n$ dimensions clearly represent center-of-mass motion in a
submanifold $\CC$, leaving a $2(n-1)$-dimensional submanifold $\RR$
describing nontrivial relative motion. We can further divide $\RR$ into a
2-dimensional piece $\SS$ which describes ``symmetric'' configurations
where the vortices are equidistant from the origin and separated from one
another by relative angles of $2\pi/n$, and a $2(n-2)$-dimensional piece
describing deviations from such symmetric configurations. The choice of
coordinates for $\SS$ can be taken to be $(\rho,\psi)$, the distance from
the origin and the angular position of a vortex relative to some reference
direction (the $x$-axis, say). The choice for $\RR\setminus\SS$ is much
less obvious, but again symmetry comes to the rescue\cite{nmon}: since
configurations in $\SS$ are symmetric under discrete rotations, any
movement starting in $\SS$ with initial velocity along $\SS$ will remain in
$\SS$, since any ``orthogonal'' motion would violate the rotational
symmetry. This implies that a choice of coordinates for $\RR$ exists for
which the induced metric is block diagonal with no mixing between $\SS$ and
$\RR\setminus\SS$. With this choice, we are free to consider just the
motion within $\SS$.

For the induced metric in $\SS$, with coordinates $(\rho,\psi)$ described
above, rotational symmetry dictates that the metric is
\begin{equation}
ds^2=f(\rho)d\rho^2+g(\rho)d\psi^2.
\label{one}
\end{equation}
Following Ruback, static configurations differing by $\psi\to\psi+2\pi/n$ are
identical, so it is tempting to impose this periodicity on the angular variable
in $\SS$. One must be cautious, however, since {\it a priori} one could arrive
at a conic singularity at the origin, which would render geodesic motion
ambiguous at that point. A straightforward generalization of Ruback's
calculation for the case of two vortices shows that near the origin the metric
for the submanifold $\SS$ is given by
\begin{equation}
ds^2\propto d\rho^2+n^2\rho^2d\psi^2,
\label{two}
\end{equation}
so that encircling the origin yields the flat-space relation between
circumference and radius: there is no conic singularity if $\psi$ and
$\psi+2\pi/n$ are identified.

Now consider geodesic motion in $\SS$ which passes through the origin. One
has initially $\psi=0$, say, and $\rho$ decreasing. This represents an
$n$-vortex head-on collision, as described above. As $\rho$ decreases to
zero, one passes to the other side of the origin in $\SS$ along a straight
line in $\SS$, and emerges with $\rho$ increasing and $\psi$ changed by
half its range, {\it i.e.}, $\psi=\pi/n$. In configuration space, this is
exactly the type of scattering by angle $\pi/n$ as seen in previous work.

This behaviour is seen outside the
Bogomol'nyi limit and beyond the geodesic approximation for vortices (until
vortex-antivortex creation becomes possible)\cite{nvort}. In constast, it is
interesting to note that for the case of baby skyrmions, {\sl fast} skyrmions
scatter at $\pi/n$, while in the geodesic approximation the scattering is by
angle zero: there is a repulsive barrier to
overcome before nontrivial scattering takes place\cite{nbaby}.

In summary, the motion of gauge theory vortices in the Bogomol'nyi limit and in
the geodesic approximation clearly shows that for $n$-vortex head-on
collisions,
the scattering is necessarily by an angle $\pi/n$, as has been seen in recent
numerical and analytical studies.

\bigskip\bigskip
I thank T. Gisiger and M. Paranjape for useful discussions, and Cairo
for inspiration.
This work was supported by the Natural Science and Engineering Research Council
of Canada.

\end{document}